\pdfoutput=1

\documentclass{INCLUDES/llncs}
\usepackage{amsfonts,amsmath}
\usepackage{graphicx}
\usepackage{url}
\usepackage{verbatim} 
\usepackage{listings}
\usepackage{amsmath}
\usepackage{subfig}
\lstnewenvironment{codex}
    {\lstset{}%
      \csname lst@SetFirstLabel\endcsname}
    {\csname lst@SaveFirstLabel\endcsname}
\lstnewenvironment{code}
    {\lstset{}%
      \csname lst@SetFirstLabel\endcsname}
    {\csname lst@SaveFirstLabel\endcsname}
\newcommand{\BI}[0]{\begin{itemize}}
\newcommand{\EI}[0]{\end{itemize}}

\newcommand{\BE}[0]{\begin{enumerate}}
\newcommand{\EE}[0]{\end{enumerate}}

\newcommand{\BX}[0]{\begin{codex}}
\newcommand{\EX}[0]{\end{codex}}
\newcommand{\BV}[0]{\begin{verbatim}}
\newcommand{\EV}[0]{\end{verbatim}}

\def \bscale1 {0.25}
\def \bscale {0.25}

\newcommand{\FIG}[4]{
\begin{figure}[htbp]
\centering
{\includegraphics[scale=#3]{./figs/#4}}
\caption{#2}
\label{#1}
\end{figure}
}

\begin{document}

\title{
  Bijective Term Encodings
}

\author{Paul Tarau}
\institute{
   {Department of Computer Science and Engineering}\\
   {University of North Texas}\\ 
%   {Denton, Texas}\\
   {\em tarau@cs.unt.edu}\\
}
\maketitle
\date{}

\label{firstpage}

\begin{abstract}
We encode/decode Prolog terms as unique natural numbers. 
 Our encodings have the following properties: 
 a) are bijective 
 b) natural numbers always decode to syntactically valid terms
 c) they work in low polynomial time in the bitsize of the representations 
 d) the bitsize of our encodings is within constant
    factor of the syntactic representation of the input.

We describe encodings of term algebras with finite signature 
as well as algorithms that separate the ``structure" of a term,
a natural number encoding of a list of balanced parenthesis,
from its ``content'', a list of atomic terms and Prolog variables.

The paper is organized as a literate Prolog program available from 
\url{http://logic.cse.unt.edu/tarau/research/2011/bijenc.pl}.
%\keywords

{\bf Keywords:}
{\em 
natural number encodings of term algebras with finite signatures
bijective G\"{o}del numberings for Prolog terms
ranking/unranking functions for tuples and lists
Catalan skeletons of Prolog terms
% bijective base-k encodings
}
\end{abstract}

\section{Introduction}

A {\em ranking/unranking} function defined on a data type is a
bijection to/from the set of natural numbers (denoted $\mathbb{N}$). 
When applied to formulas or proofs, ranking functions are usually called
{\em G\"{o}del numberings} as they have originated in arithmetization
techniques used in the proof of G\"{o}del's incompleteness results
\cite{Goedel:31,conf/icalp/HartmanisB74}. 
In  G\"{o}del's original encoding \cite{Goedel:31}, given that
primitive operation and variable symbols in a formula are mapped to exponents of 
distinct prime numbers, factoring is required for
decoding, which is therefore intractable for formulas of non-trivial size. 
As this mapping is not a surjection, there are codes that decode
to syntactically invalid formulas. This key difference  also applies to
alternative G\"{o}del numbering schemes (like G\"odel's beta-function), while
ranking/unranking functions, as used in combinatorics, are bijective mappings.

Besides codes associated to formulas,
a wide diversity of common computer operations, ranging
from data compression and serialization to
data transmissions and cryptographic codes
are essentially bijective encodings between 
data types. They provide a variety of services ranging from
free iterators and random objects to data compression and succinct
representations. Tasks like serialization and persistence are facilitated
by simplification of reading or writing operations without the need of
special-purpose parsers.

The main focus of this paper is designing an efficient bijective G\"odel
numbering scheme (i.e. a ranking/unranking bijection) for {\em term algebras},
essential building blocks for various data types and programming language constructs.

The resulting G\"odel numbering algorithm, the main contribution of the paper,
enjoys the following properties:
\begin{enumerate}
\item the mapping is bijective 
\item natural numbers always decode to syntactically valid terms
 \item it works in time low polynomial in the bitsize of the representations
 \item the bitsize of our encoding is within
 constant factor of the syntactic representation of the input.
\end{enumerate}

These properties ensure that our algorithm can be applied to derive compact
serialized representations for various formal systems and programming language
constructs.

\section{Tuple Encodings} \label{tuple}
We will now define a few primitive operations in terms of a small
set of bitwise primitives with known asymptotic complexity.
Assuming a copying implementation of arbitrary size integers
each of the following operations are at most linear in the bitsize
of their operand {\tt N} and some can be considered constant time
when this operand fits in a machine word as well as when
an efficient mutable implementation of arbitrary length integers is used.
\begin{code}
first_bit(N,Bit):- Bit is 1 /\ N.
times_exp2(N,K,R):-R is N << K.
div_by_exp2(N,K,R):-R is N >> K.
predecessor(N,R):-R is N-1.
successor(N,R):-R is N+1.
\end{code}

First we define the {\tt k\_deflate} and {\tt k\_inflate} operations.
{\tt k\_deflate} can be seen as collecting each k-th bit 
from a number's binary representation and aggregates the result 
into a new natural number. {\tt k\_inflate} can be seen as building
a new natural number by inserting {\tt 0}s in every position
except in each k-th position where the bits of its argument {\tt N}
are placed. However, we avoid direct bitlist manipulation by
expressing them in terms of the previously defined
arbitrary length integer operations.

\begin{code}
k_deflate(_,0,0).
k_deflate(K,N,R):-N>0,
  div_by_exp2(N,K,A),
  k_deflate(K,A,B),
  times_exp2(B,1,C),
  first_bit(N,D),
  R is C\/D.
\end{code}
\begin{code}
k_inflate(_,0,0).
k_inflate(K,N,R):-N>0,
  div_by_exp2(N,1,A),
  k_inflate(K,A,B),
  times_exp2(B,K,C),
  first_bit(N,D),
  R is C\/D.
\end{code}
The following example illustrates their use:
{\small \begin{verbatim}
?- k_inflate(3,42,X),k_deflate(3,X,Y).
X = 33288,
Y = 42 .
\end{verbatim}}

We can define a bijective decomposition of a natural number {\tt N}
as a tuple of {\tt K} natural numbers in terms of {\tt k\_deflate}
and our primitive bitwise operations.

The function {\tt to\_tuple:} $Nat \rightarrow Nat^k$ converts a natural 
number to a $k$-tuple by splitting its bit representation into $k$ groups, 
from which the $k$ members in the tuple are finally rebuilt. This operation 
can be seen as a transposition of a bit matrix obtained by expanding 
the number in base $2^k$:
\begin{code}
to_tuple(K,N,Ns):-K>0,
  predecessor(K,K1),
  numlist(0,K1,Ks),
  maplist(div_by_exp2(N),Ks,Ys),
  maplist(k_deflate(K),Ys,Ns).
\end{code}
Note the use of the SWI-Prolog library predicates {\tt numlist} that generates
a list of integers in increasing order and {\tt maplist} that applies a closure to
lists of corresponding arguments.
To convert a $k$-tuple back to a natural number we will merge their 
bits, $k$ at a time. This operation can be seen as the transposition of a bit 
matrix obtained from the tuple, seen as a number in base $2^k$, but we
implement it more efficiently in terms of bitwise operations
on integers. Note the use of the SWI-Prolog library predicate {\tt sumlist}
that computes the sum of a list of numbers.
\begin{code}
from_tuple(Ns,N):-
  length(Ns,K),K>0,
  predecessor(K,K1),
  maplist(k_inflate(K),Ns,Xs),
  numlist(0,K1,Ks),
  maplist(times_exp2,Xs,Ks,Ys),
  sumlist(Ys,N).
\end{code}
The following example shows the mapping of {\tt 42} to a $3$-tuple and the encoding 
back to {\tt 42}.
{\small \begin{verbatim}
?- to_tuple(3,42,T),from_tuple(T,N).
T = [2, 1, 2],
N = 42 .
\end{verbatim}}
Fig. \ref{isot42} shows multiple steps of the same decomposition, 
with shared nodes collected in a DAG. Note that 
markers on edges indicate argument positions.
\FIG{isot42}{42 after repeated 3-tuple expansions}{0.50}{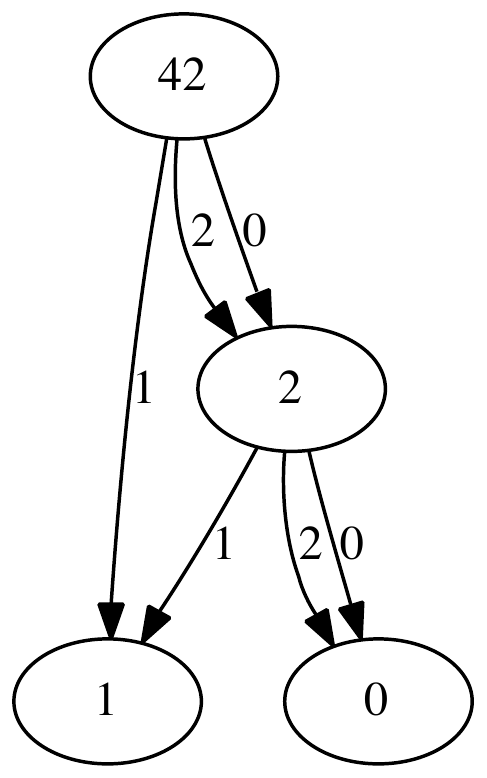}
%\HFIGS{3tuple}{Repeated 3-tuple expansions}{42}{2008}{42tuple}{2008tuple}

Note that one can now define {\em pairing functions}, i.e. {\em bijections between
natural numbers and pairs of natural numbers},
as specializations of the tupling/untupling
predicates:
\begin{code}
to_pair(N,A,B):-to_tuple(2,N,[A,B]).
\end{code}
\begin{code}
from_pair(X,Y,Z):-from_tuple([X,Y],Z).
\end{code}
One can observe that {\tt to\_pair} and {\tt from\_pair} are the same as the
functions defined in Steven Pigeon's PhD thesis 
on Data Compression \cite{pigeon} and also known as Morton-codes
with uses in indexing of spatial databases \cite{lawder99}.

\section{G\"odel numberings of a term algebra with finite signature}

Traditionally a term algebra is defined over a 
finite set of functions symbols of given arities. 
Constants can be singled out as a special set
or considered function symbols of arity 0.
Term algebras are 
{\em free magmas\footnote{See \url{http://wikipedia.org/wiki/Free_object}}}
induced by a set of variables and a
set of function symbols of various arities (0 included), called {\tt signature}, that
are closed under the operation of inserting terms as arguments of function
symbols. In various logic formalisms a term algebra is called a Herbrand
Universe.

Having a bijective encoding over a signature and a (finite) set  of 
variables (seen as input ``wires") is also useful for synthesizing 
code over a set of functions - 
for instance a library of 
logic gates, in the case of circuit synthesis, as well as for 
generating random terms
of a given signature for testing purposes.

Besides being bijective, it is useful if the mapping
relates terms to numbers of comparable representation size
and if it works in linear or low polynomial time to
be useful for practical applications.
 
We will denote {\tt Vs, CSyms, FSyms} the sets of variables, constant symbols and
function symbol/arity pairs, respectively, that parameterize 
the converter from terms to codes {\tt term2nat/5} and the converter
from codes to terms, {\tt nat2term/5}.

Given that {\tt Vs, CSyms} are finite, we map them bijectively to
the ranges {\tt [0..LV-1]} (variables), {\tt [LV..LV+LC-1]} (constants).
The predicate {\tt term2nat} precomputes these values and
then calls the recursive converter {\tt t2n}.

\begin{code}
term2nat(Vs,CSyms,FSyms,T, X):-
  length(CSyms,LC),
  length(FSyms,LF),
  length(Vs,LV),
  LVC is LV+LC,
  LVC>0,
  t2n(LV,LC,LF,LVC,Vs,CSyms,FSyms,T, X).
\end{code}
The predicate {\tt t2n} uses {\tt lookup\_var} and the built-in {\tt nth0/3}
to look-up indices associated to
variable, constant and function symbols.
For compound terms, these values are combined with
values computed recursively on their arguments
and then merged using {\tt from\_tuple} into 
natural numbers. 

Note that the index {\tt L}
of the function symbol {\tt F/K} computed by {\tt nth0/3}
is multiplied with the length {\tt LF} of
the list of function
symbols {\tt FSyms}. This operation will
be reversed using modulo and quotient
when converting back.
\begin{code}
t2n(LV,_LC,_LF,_LVC,Vs,_CSyms,_FSyms,V, X):-var(V),!,
  lookup_var(I,Vs,V),
  I>=0,I<LV,X=I.
t2n(LV,_LC,_LF,_LVC,_Vs,CSyms,_FSyms,C, X):-atomic(C),!,
  nth0(I,CSyms,C),
  X is I+LV.
t2n(LV,LC,LF,LVC,Vs,CSyms,FSyms,T, X):-compound(T),
  T=..[F|Ts],
  nth0(L,FSyms,F/K),
  K>0,
  length(Args,K),
  P=..[t2n,LV,LC,LF,LVC,Vs,CSyms,FSyms],
  maplist(P,Ts,Args),
  from_tuple(Args,N),
  X is LVC+LF*N+L.
\end{code}
\begin{code}
lookup_var(N,Xs,X):-lookup_var(X,Xs,0,N).
\end{code}
\begin{code}
lookup_var(X,[Y|_],N,N):-X==Y.
lookup_var(X,[_|Xs],N1,N3):-
  N2 is N1+1,
  lookup_var(X,Xs,N2,N3).
\end{code}
The predicate {\tt nat2term} reverses the process,
using the same lists {\tt Vs,CSyms,FSyms} to map
variables, constants and function symbols to
natural number codes, by calling the recursive
converter {\tt n2t/8}.
\begin{code}
nat2term(Vs,CSyms,FSyms,X, T):-X>=0,
  length(CSyms,LC),
  length(FSyms,LF),
  length(Vs,LV),
  LVC is LV+LC,LVC>0,
  n2t(LV,LC,LF,LVC,Vs,CSyms,FSyms,X, T).
\end{code}
The recursive converter {\tt n2t} uses dictionaries
{\tt Vs, CSyms, FSyms} to map natural numbers
to the corresponding, functions, constant and variable
terms, uniformly. Note the use of the library
predicate {\tt nth0} that associates an index, starting at 0,
to a term on a list. It also uses Prolog's {\tt univ} to
build a closure {\tt P} that with help from 
{\tt maplist} applies it recursively.
\begin{code}
n2t(LV,_LC,_LF,_LVC,Vs,_CSyms,_FSyms,X, V):-X<LV,!,
  nth0(X,Vs,V).
n2t(LV,_LC,_LF,LVC,_Vs,CSyms,_FSyms,X, C):-LV=<X,X<LVC,!,
  X0 is X-LV,
  nth0(X0,CSyms,C).
n2t(LV,LC,LF,LVC,Vs,CSyms,FSyms,X, T):-X>=LVC,
  X0 is X-LVC,
  N is X0 // LF,
  L is X0 mod LF,
  nth0(L,FSyms,F/K),
  K>0,
  to_tuple(K,N,Args),
  P=..[n2t,LV,LC,LF,LVC,Vs,CSyms,FSyms],
  maplist(P,Args,Ts),
  T=..[F|Ts].
\end{code}
Note the use of the predicate {\tt to\_tuple} with
length {\tt K} based on the arity of each function symbol,
which splits the natural number {\tt N}
in a list of codes {\tt Args} to be used
recursively to build the subterms associated
to the function symbol {\tt F/K}.

A first example shows that starting from a term {\tt T} we obtain a
natural number from which the same term {\tt T} is recovered. 
Note that the two side of the transformer are parameterized 
by the same lists of variables, constants and function symbols.
{\small \begin{verbatim}
?- T=f(a,f(X,g(Y))),Vs=[X,Y],Cs=[a],Fs=[f/2,g/1],
   term2nat(Vs,Cs,Fs,T,N),nat2term(Vs,Cs,Fs,N,T_again).
T = f(a, f(X, g(Y))),
Vs = [X, Y],
Cs = [a],
Fs = [f/2, g/1],
N = 17439,
T_again = f(a, f(X, g(Y))) .
\end{verbatim}}

The next example shows that starting from any natural number e.g. {\tt 2012}
we obtain a term that in turn is converted back to the same number.
{\small \begin{verbatim}
?- N=2012,Vs=[X,Y],Cs=[a,b],Fs=[f/2,g/1],
   nat2term(Vs,Cs,Fs,N,T),term2nat(Vs,Cs,Fs,T,N_again).
N = 2012,
Vs = [X, Y],
Cs = [a, b],
Fs = [f/2, g/1],
T = f(f(Y, b), f(b, a)),
N_again = 2012 .
\end{verbatim}}

Finally, the following example (where {\tt '->'} is seen as logical implication),
hints towards an application to circuit synthesis. When combined with a fast
bitstring-based boolean evaluator (see \cite{iclp07}) terms associated
with natural numbers can be tried out to see if the result
of their boolean evaluation matches a given specification.
{\small \begin{verbatim}
?- N=2012,Vs=[A,B],Cs=[0],Fs=['->'/2],
   nat2term(Vs,Cs,Fs,N,T),term2nat(Vs,Cs,Fs,T,N_again).
N = 2012,
Vs = [A, B],
Cs = [0],
Fs = [ (->)/2],
T = (((B->A)->0->A)-> (0->A)->B),
N_again = 2012 .
\end{verbatim}}

\begin{comment}
\begin{code}
% number to term test
t(N,Vs-T):-Vs=[_,_],nat2term(Vs,[0],['->'/2],N,T).

% term to number test
n(Vs-T,N):-term2nat(Vs,[0],['->'/2],T,N).

% number to term test
t1(N,Vs-T):-length(Vs,3),nat2term(Vs,[0,1],['+'/2,'*'/2],N,T).

% term to number test
n1(Vs-T,N):-term2nat(Vs,[0,1],['+'/2,'*'/2],T,N).

% number to term test
t2(N,Vs-T):-length(Vs,2),nat2term(Vs,[a,b,c],['f'/2,'g'/3,'h'/1],N,T).

% term to number test
n2(Vs-T,N):-term2nat(Vs,[a,b,c],['f'/2,'g'/3,'h'/1],T,N).

test:-between(0,100,I),t(I,X),n(X,N),portray_clause([I=N,X]),fail.

\end{code}
\end{comment}

One can generate random terms with a given signature based on a
natural number of a given bitsize as follows.
\begin{code}
ranterm(Bits,Vs,Cs,Fs, T):-
  N is random(2^Bits),
  nat2term(Vs,Cs,Fs,N,T).
\end{code}
This can be useful in generating random 
arithmetic expressions or boolean functions for
testing purposes.
{\small \begin{verbatim}
?- Vs=[A,B,C],ranterm(100,Vs,[],['+'/2,'*'/2],T).
Vs = [A, B, C],
T = (B+ (C+A))* ((A+B)*A* ((A+A)*C))+ (A+B+A*A+ (B+ (A+A))*(B+A))+ 
    (B*A*B* (B+B)* (C+ (C+B))+ (A* (A+A)+C*A)* ((B+A)* ((A+A)*C))) .

?- Vs=[A,B,C,D],ranterm(50,Vs,[0,1],[and/2,or/2,not/1],T).
Vs = [A, B, C, D],
T = and(not(not(or(or(not(0), A), or(and(B, B), A)))), 
    or(or(and(A, A), or(D, A)), not(or(C, not(B))))) .
\end{verbatim}}

\section{Bijective encodings of Prolog atoms}
Prolog provides a mapping between its symbols and their
character codes. To obtain an encoding of strings linear in their bitsize
we need a general mechanism to map arbitrary combinations
of {\tt k} symbols to natural numbers. 

\subsection{Encoding numbers in bijective base-k} \label{bijnat}

The conventional numbering system does not provide a bijection between
arbitrary combinations of digits and natural numbers, given that leading 0s are
ignored. For this purpose we need to use {\em numbers in bijective
base-k}\footnote{We refer to
\url{http://en.wikipedia.org/wiki/Bijective_numeration} for the
historical origins of the concept and the 
properties of this number representation.}. 
First we start with the mapping from list of digits in {\tt [0..k-1]}
to a natural number defined by the predicate {\tt from\_bbase/3}
\begin{code}
from_bbase(Base,Xs,R):-
  maplist(successor,Xs,Xs1),
  from_base1(Base,Xs1,R).
\end{code}
\begin{code}
from_base1(_Base,[],0).
from_base1(Base,[X|Xs],R):-X>0,X=<Base,
  from_base1(Base,Xs,R1),
  R is X+Base*R1.
\end{code}

\begin{code}   
to_bbase(Base,N,Xs):-
  to_base1(Base,N,Xs1),
  maplist(predecessor,Xs1,Xs).

to_base1(_,0,[]).
to_base1(Base,N,[D1|Ds]):-N>0,
   Q is N//Base,
   D is N mod Base,
   (D==0->D1=Base;D1=D),
   (D==0->Q1 is Q-1;Q1=Q),
   (Q1==0->Ds=[];to_base1(Base,Q1,Ds)).
\end{code}
Note that the predicates {\tt from\_bbase} and {\tt to\_bbase} are
parametrized by the base of numeration which
should be the same when encoding and decoding.
{\small \begin{verbatim}
?- to_bbase(7,2012,Ds),from_bbase(7,Ds,N).
Ds = [2, 6, 4, 4],
N = 2012 .
\end{verbatim}}
This encoding will turn out to be useful for
symbols of a finite alphabet.

\subsection{Encoding strings}
Strings can be seen just as a notational equivalent
of lists of natural numbers written in bijective base-$k$.
For simplicity (and to avoid unprintable characters as a result of applying the
inverse mapping) we will assume that our strings naming 
functions are built only using lower case ASCII characters.
\begin{code}
c0(A):-[A]="a".
c1(Z):-[Z]="z".

base(B):-c0(A),c1(Z),B is 1+Z-A.
\end{code}
Next, we define the bijective base-k encodings
\begin{code}
string2nat(Cs,N):-
  base(B), 
  maplist(chr2ord,Cs,Ns),
  from_bbase(B,Ns,N).
\end{code}
\begin{code}
nat2string(N,Cs):-N >= 0,
  base(B),
  to_bbase(B,N,Xs),
  maplist(ord2chr,Xs,Cs).
\end{code}
\begin{code}  
chr2ord(C,O):-c0(A),C>=A,c1(Z),C=<Z,O is C-A.
ord2chr(O,C):-O>=0,base(B),O<B,c0(A),C is A+O.
\end{code}
We obtain an encoder for strings working as follows:
{\small \begin{verbatim}
?- Cs="hello",string2nat(Cs,N),nat2string(N,CsAgain).
Cs = [104, 101, 108, 108, 111],
N = 7073802,
CsAgain = [104, 101, 108, 108, 111] .

?- nat2string(2012,Cs),string2nat(Cs,N).
Cs = [106, 121, 98],
N = 2012 .
\end{verbatim}}
And finally we can obtain a bijective encoding of Prolog atoms as
\begin{code}
atom2nat(Atom,Nat):-atom_codes(Atom,Cs),string2nat(Cs,Nat).

nat2atom(Nat,Atom):-nat2string(Nat,Cs),atom_codes(Atom,Cs).
\end{code}

\section{``Catalan skeletons" of Prolog terms} \label{termenc}

We will now turn to encodings focusing on the separation 
of the structure and the content of Prolog terms.
The connection between balanced parenthesis languages and
a large number of different data types (among which we find
multi-way and binary trees)
in the {\em Catalan family}
is known to combinatorialists
\cite{berstel2002formal,liebehenschel2000ranking}.
We will start by mapping a term to a ``skeleton"
representing its structure as a list of
balanced parentheses.

\subsection{An injective-only structure encoding}

We sketch here an encoding mechanism that might also be useful to Prolog
implementors interested in designing alternative heap representations for new
Prolog runtime systems or abstract machine architectures as well as hashing mechanisms
for ground terms or variant checking for tabling.

First we provide an encoding that separates the ``structure" of a term {\tt T},
expressed as a balanced parenthesis 
language\footnote{A member of the {\em Catalan family} of combinatorial objects.} 
representation 
{\tt Ps} and a list of atomic terms and Prolog variables {\tt As},
seen as a symbol table that stores the ``content" of the terms:
\begin{code}
term2bitpars(T,[0,1],[T]):-var(T).
term2bitpars(T,[0,1],[T]):-atomic(T).
term2bitpars(T,Ps,As):-compound(T),term2bitpars(T,Ps,[],As,[]).
\end{code}
\begin{code}
term2bitpars(T,Ps,Ps)-->{var(T)},[T].
term2bitpars(T,Ps,Ps)-->{atomic(T)},[T].
term2bitpars(T,[0|Ps],NewPs)-->{compound(T),T=..Xs},
  args2bitpars(Xs,Ps,NewPs).
\end{code}
\begin{code}  
args2bitpars([],[1|Ps],Ps)-->[].
args2bitpars([X|Xs],[0|Ps],NewPs)-->
  term2bitpars(X,Ps,[1|XPs]),
  args2bitpars(Xs,XPs,NewPs).  
\end{code}  
The encoding is reversible, i.e. the term T can be recovered:
\begin{code}
bitpars2term([0,1],[T],T).
bitpars2term([P,Q,R|Ps],As,T):-bitpars2term(T,[P,Q,R|Ps],[],As,[]).
\end{code}
\begin{code}
bitpars2term(T,Ps,Ps)-->[T].
bitpars2term(T,[0|Ps],NewPs)-->
  bitpars2args(Xs,Ps,NewPs),{T=..Xs}.
\end{code}
\begin{code}  
bitpars2args([],[1|Ps],Ps)-->[].
bitpars2args([X|Xs],[0|Ps],NewPs)-->
  bitpars2term(X,Ps,[1|XPs]),
  bitpars2args(Xs,XPs,NewPs).  
\end{code}
The two transformations work as follows:
{\small \begin{verbatim}
?- term2bitpars(f(g(a,X),X,42),Ps,As),
   bitpars2term(Ps,As,T).
Ps = [0,0,1,0,0,0,1,0,1,0,1,1,1,0,1,0,1,1],
As = [f,g,a,X,X,42],
T=f(g(a,X),X,42) .
\end{verbatim}}
By using this encoding one can further aggregate bitlists into 
natural numbers with {\tt term2inj\_code}
by converting the resulting bitlists seen as 
bijective-base 2 digits and then convert 
them back with {\tt inj\_code2term}.
\begin{code}
term2inj_code(T,N,As):-
  term2bitpars(T,Ps,As),
  from_bbase(2,Ps,N).
\end{code}
\begin{code}  
inj_code2term(N,As,T):-
  to_bbase(2,N,Ps),
  bitpars2term(Ps,As,T).
\end{code}
working as follows:
{\small \begin{verbatim}    
?- term2inj_code(f(a,g(X,Y),g(Y,X)),N,As),inj_code2term(N,As,T).
N = 131364115,
As = [f, a, g, X, Y, g, Y, X],
T = f(a, g(X, Y), g(Y, X)) .
\end{verbatim}}

\begin{comment}
One can complete the encoding by hashing the symbol table into a
list of small integers that can be encoded as a natural number
and then aggregated with the result of {\tt term2inj\_code} using a pairing function.
\end{comment}
Note however that this encoding is injective only i.e. not every natural number
is a code of a term.

We will next describe a bijective encoding to ``Catalan skeletons" 
which abstract away the structure of a Prolog term as a unique natural
number code. 

\subsection{A Diophantine decomposition of natural numbers}
First, we need a mechanism to bijectively encode/decode 
the actual information content of term as well as
the arities associated to its function symbols.

As an immediate consequence of the unique decomposition
of natural numbers in prime factors,
the Diophantine equation
\begin{equation}
2^x(2y+1)=z
\end{equation}
has, for any positive natural number {\tt z} a unique solution {\tt  (x,y)}.

Using the {\tt lsb/1} function that returns the least significant bit of a 
natural number (available, for instance, in SWI-Prolog
and easy to emulate in other Prologs) one can define:
\begin{code}
cons(X,Y, Z):-Z is ((Y<<1)+1)<<X.

decons(Z, X,Y):-Z>0, X is lsb(Z), Y is Z>>(X+1).
\end{code}
We will use these predicates to decompose a natural 
number {\tt Z>0} into {\tt X} and {\tt Y} such that
{\tt X} is well suited to work as the length of a tuple
and {\tt Y} to provide the members of a tuple of length {\tt X},
in a reversible way.

\subsection{A bijection between natural numbers and lists}

By combining {\tt cons/3} and {\tt decons/3} (which aggregate/separate ``length'' and ``content'')
with {\tt to\_tuple} and
{\tt from\_tuple} (which aggregate/separate a ``content'' of fixed ``length''),
we obtain an bijection between lists of numbers
and numbers of size proportional to the bit representations 
of the operands.
\begin{code}
nat2nats(0,[]).
nat2nats(N,Ns):-N>0,
  decons(N,L1,N1),
  L is L1+1,
  to_tuple(L,N1,Ns).
\end{code}
\begin{code}  
nats2nat([],0).
nats2nat(Ns,N):-
  length(Ns,L),
  L1 is L-1,
  from_tuple(Ns,N1),
  cons(L1,N1,N).
\end{code}
The following example illustrates that this encoding is a bijection:
{\small \begin{verbatim}
?- nat2nats(2012,Ns),nats2nat(Ns,N).
Ns = [7, 7, 2],
N = 2012 .
\end{verbatim}}

\subsection{A bijection between natural numbers and lists of balanced parenthesis}
We can build a bijection between lists of balanced parenthesis and
natural numbers
by encoding sublists, recursively with {\tt nats2nat/1}
while parsing them with a DCG.
\begin{code}
pars2nat(Xs,T):-pars2nat(0,1,T,Xs,[]).
\end{code}
\begin{code}
pars2nat(L,R,N) --> [L],pars2nats(L,R,Xs),{nats2nat(Xs,N)}.

pars2nats(_,R,[]) --> [R].
pars2nats(L,R,[X|Xs])-->pars2nat(L,R,X),pars2nats(L,R,Xs).
\end{code}
The inverse mapping works in a similar way, using {\tt nat2nats}
to recursively generate the lists of balanced parenthesis using
a DCG.
\begin{code}
nat2pars(N,Xs):-nat2pars(0,1,N,Xs,[]).
\end{code}
\begin{code}
nat2pars(L,R,N) --> {nat2nats(N,Xs)},[L],nats2pars(L,R,Xs).

nats2pars(_,R,[]) --> [R].
nats2pars(L,R,[X|Xs])-->nat2pars(L,R,X),nats2pars(L,R,Xs).
\end{code}

The following example illustrates that the two mappings
are indeed invertible.
{\small \begin{verbatim}
?- nat2pars(2012,Ps),pars2nat(Ps,N).
Ps = [0,0,0,0,0,1,1,1,1,0,0,0,0,1,1,1,1,0,0,1,0,1,1,1],
N = 2012 .
\end{verbatim}}

\subsection{Bijective Catalan skeletons of Prolog terms}

By combing the converters between terms to lists of parenthesis
with a bijection provided by {\tt pars2nat} and {nat2pars}
we obtain:
\begin{code}
term2code(T,N,As):-
  term2bitpars(T,Ps,As),
  pars2nat(Ps,N).
\end{code}
\begin{code}
code2term(N,As,T):-
  nat2pars(N,Ps),
  bitpars2term(Ps,As,T).
\end{code}  

As as the following example shows the encoding is indeed reversible:
{\small \begin{verbatim}
?- term2code(f(a,g(X,Y),g(Y,X)),N,As),code2term(N,As,T).
N = 786632,
As = [f, a, g, X, Y, g, Y, X],
T = f(a, g(X, Y), g(Y, X)) .
\end{verbatim}}
Not also the succinctness, by comparison to the usual Prolog heap representations,
of the ``Catalan structure" of the term.

\section{Related work} \label{related}

This paper can be seen as an application to the data transformation framework 
\cite{ppdp09pISO} which helps gluing together the pieces
needed for the derivation of our bijective encoding of term algebras,
including algebras with finite signatures,
the novel contribution of this paper.

We have not found in the literature an encoding scheme
for term algebras that is {\em bijective}, nor an encoding that
is computable both ways with effort proportional to the
size of the inputs.

On the other hand, {\em ranking} functions for sequences 
can be traced back to G\"{o}del numberings
\cite{Goedel:31,conf/icalp/HartmanisB74} associated to formulas. 
Together with their inverse {\em unranking} functions they are also 
used in combinatorial generation
algorithms
\cite{conf/mfcs/MartinezM03,knuth_trees}.
Pairing functions have been used in work on decision problems as early as
\cite{robinson50}. A
typical use in the foundations of mathematics is
\cite{DBLP:journals/tcs/CegielskiR01}.
An extensive study of various pairing functions and their 
computational properties is presented in 
\cite{DBLP:conf/ipps/Rosenberg02a}.

The closest reference on encapsulating bijections
as a programming language data type is \cite{bijarrows} 
and Conal Elliott's composable
bijections Haskell module \cite{bijeliot}.
\cite{Kahl-Schmidt-2000a} uses a similar category theory inspired framework
implementing relational algebra, also in a Haskell setting.

\section{Conclusion} \label{concl}
We have described a compact bijective G\"odel numbering scheme for term
algebras. The algorithm can be made to work in linear time and has applications
ranging from generation of random instances to exchanges of structured data between
declarative languages and/or theorem provers and proof assistants. 
We foresee some practical
applications as a generalized serialization mechanism
usable to encode complex
information streams with heterogeneous subcomponents - 
for instance as a mechanism for
sending serialized objects over a network.
Also, given that our encodings are {\em bijective}, they can be used to generate
random terms, which in turn, can be used to represent
random code fragments. This could have applications ranging from
generation of random tests to representation
of populations in genetic programming.

\section*{Acknowledgment} We thank NSF (research grant 1018172) for support.

\bibliographystyle{INCLUDES/splncs}
\bibliography{INCLUDES/theory,tarau,INCLUDES/proglang,INCLUDES/biblio,INCLUDES/syn}

\end{document}